\documentclass[useAMS,usenatbib]{mn2e}
\usepackage{natbib}

\newcommand*{\Msun}{\ensuremath{\mathrm{M_\odot}}}%

\usepackage{graphicx}
\usepackage{epstopdf}
\title[Star Formation in XMMU J2235.3-2557 at $z=1.4$]{Star Formation in the XMMU J2235.3-2557 Galaxy Cluster at $z=1.39$}

\author[Bauer et al.]{Amanda E.~Bauer$^{1}$\thanks{email: amanda.bauer@nottingham.ac.uk},
Ruth~Gr\"utzbauch$^{1}$\thanks{email: ruth.grutzbauch@nottingham.ac.uk}, Inger J\o rgensen$^{2}$\thanks{email: Inger@gemini.edu}, Jesus Varela$^{3,4}$\thanks{email: jmvarela@iac.es}, \newauthor Marcel Bergmann\thanks{email: marcel.bergmann@gmail.com}\\
$^{1}$ School of Physics and Astronomy, University of Nottingham, UK\\
$^{2}$ Gemini Observatory, 670 N. A'ohoku Pl., Hilo, HI 96720 \\
$^{3}$ Instituto de Astrof\'isica de Canarias (IAC), E-38200 La Laguna, Tenerife, Spain\\
$^{4}$ Depto. Astrof\'isica, Universidad de La Laguna (ULL), E-38206 La Laguna, Tenerife, Spain\\
}

\begin{document}

\date{Accepted -- . Received --}

\maketitle

\begin{abstract}
We present the first results of a  narrow-band photometric study of the massive galaxy cluster XMMU J2235.3-2557 at $z=1.39$. We obtained deep $H$ narrow-band imaging with NIRI on Gemini North, corresponding to H$\alpha$ emission at the cluster's redshift.  Our sample consists of 82 galaxies within a radius of $\sim$500 kpc, ten of which are spectroscopically confirmed cluster members.  Sixteen galaxies are identified as excess line-emitters.  Among just the excess line-emitting galaxies we find an average SFR of 3.6$~\pm~1.3$~M$_\odot$yr$^{-1}$.  For spectroscopically confirmed cluster members we find a correlation between $H$ broad-band magnitude and SFR such that brighter galaxies have lower SFRs. The probability that SFR and magnitude of confirmed members are uncorrelated is 0.7\%.  We also find a correlation between SFR and distance from the cluster centre for both confirmed and excess line-emitting candidate members, with a probability of 5\% for there to be no correlation among confirmed members.  All excess line-emitting candidate cluster members are located outside a radius of 200 kpc. We conclude that star formation is effectively shut off within the central 200 kpc radius ($R_{QUENCH}~\sim~200$~kpc) of this massive galaxy cluster at $z=1.39$, when the universe was only 4.5 Gyr old. 
\end{abstract}


\section{Introduction}\label{sec:intro}

Galaxy evolution and formation are dependent on stellar mass and environment.  The build up of stellar mass in the early universe occurred rapidly, resulting in massive galaxies with old stellar populations detected by $z = 2$.  The global star formation history of the universe indicates that as star formation rates (SFRs) in high mass galaxies decreased steeply from $z=2$ to 1, the most active sites of star formation shifted to progressively lower mass galaxies over time \citep{Cowie96,BE00,Bauer05,Juneau2005}.  The environmental effect is that galaxies in high-density regions tend to be older and form stars on shorter time scales than those in lower-density regions \citep{Butcher84,Dressler97,Poggianti99,Thomas05}.  Therefore the dense regions of distant galaxy clusters are ideal environments to study the first galaxies in the universe to stop forming stars.   

Studies of clusters at intermediate redshifts have found strong evidence that star formation has shut off in the central regions.  Observations of the well-studied galaxy cluster RXJ0152.7-1357 at $z=0.83$ show that galaxies towards the centre are redder and harbour older stellar populations \citep{Jorgensen2005,Demarco2005,Tanaka2005,Marcillac2007,Patel2009} than galaxies in the lower density outskirts.  A similar result was found by \citet{Koyama09} for the RXJ1716.4+6708 cluster at $z=0.81$, although a different technique was employed.  \citet{Koyama09} used narrow-band H$\alpha$ imaging and found no H$\alpha$ emitters within a radius of 0.25 Mpc of the cluster core.  \citet{Finn05} also use H$\alpha$ to study three clusters at $z=0.7-0.8$.  They find that two of the three clusters show a decrease in H$\alpha$ fraction towards cluster centre, indicating that star formation has largely shut off inside the core of these intermediate redshift clusters.  

Recently, studies of high redshift clusters have investigated the presence of a relation between star formation and the distance from the cluster centre.  In the RDCS J1252.9-2927 cluster at $z = 1.24$, \citet{Tanaka2009} find a population of star-forming galaxies at the outskirts of the cluster, which are absent from the cluster core.  These results differ from the work of \citet{Hayashi10} who observe one of the most distant clusters known, XMMXCS J2215.9-1738 at $z=1.46$, and find no evidence for decreasing SFRs towards the cluster core.  The period of $z=1.0-1.5$ is crucial for understanding the shut down of star formation in cluster galaxies.    

In this study, we investigate galaxies inside the most massive distant galaxy cluster yet detected, XMMU J2235.3-2557  (``XMMU2235'') at $z=1.393$.   This distant galaxy cluster was discovered serendipitously by the orbiting XMM-Newton X-ray telescope by \citet{Mullis05} and remains one of the highest redshift clusters known.   \citet{Rosati2009} confirm cluster membership of 34 individual galaxies in XMMU2235 based on spectroscopic data.  Different methods have been used to estimate the total mass within 1 Mpc for XMMU2235.  \citet{Jee2009} estimate the projected mass to be $(8.5\pm 1.7) \times 10^{14}\Msun$ from a weak lensing study, which is consistent  with the projected X-ray mass calculated by \citet{Rosati2009} who find M$_{tot,proj}$~=~$(9.3~\pm~2.1)~\times~10^{14}\Msun$, making this the most massive galaxy cluster known at $z>1$.

\citet{Lidman2008} created a near-IR colour-magnitude diagram of XMMU2235 using VLT observations and found that galaxies detected in the core of the cluster are red, while galaxies surrounding the core show much more diverse properties.  They also use simple stellar population models to determine that galaxies in the core of XMMU2235 are already old, with typical ages of 3 Gyr.  \citet{Rosati2009} examine spectra taken with VLT of 16 spectroscopically-identified galaxies that have no [OII] emission.  They stack the spectra into two bins based on the distance from the cluster centre, then fit SED models to derive star formation histories.  They find that galaxies in the core are consistent with having experienced a single burst of star formation at $z=5.3$, while passive galaxies beyond 200 kpc of the cluster centre show signatures of more extended periods of star formation, lasting possibly to redshifts of $z=2$.  
 
In this paper, we present  narrow-band $H$ (1.57$\mu$m) observations of XMMU2235, which correspond to the wavelength region where galaxies in this cluster emit H$\alpha$.  We measure for the first time individual SFRs for 82 galaxies in XMMU2235 out to a cluster radius of 500 kpc.  In Section~$\ref{sec:data}$, we describe the near-infrared observations of the cluster, data reduction, and the calculation of star formation rates.  We discuss our results in Section~$\ref{sec:results}$, present a discussion in Section~$\ref{sec:discussion}$, and end with a summary and conclusions in Section~\ref{sec:summ}.    

Throughout the paper we assume the standard $\Lambda$CDM cosmology, a flat universe with $\Omega_\Lambda = 0.73$,  $\Omega_M = 0.27$ and a Hubble constant of $H_0 = 72 $ km s$^{-1}$ Mpc$^{-1}$.

\begin{figure}
\includegraphics[width=0.36\textwidth, angle=270]{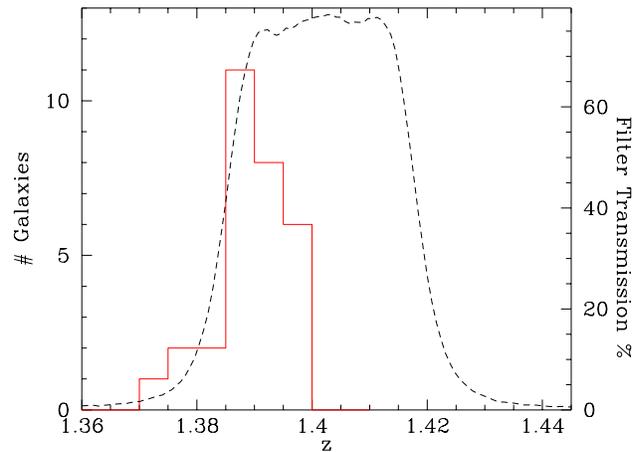}
\caption{Transmission curve of the $\lambda 1.57 \mu$m narrow-band filter (dotted line), projected to show redshift coverage, overplotted with the redshift histogram (red line) of known spectroscopic cluster members from \citet{Rosati2009}. \label{fig:filter}}
\end{figure}


\section{Observations and data reduction}\label{sec:data}

\subsection{Observations}\label{sec:observations}

We obtained observations of the XMMU2235 galaxy cluster with the Near InfraRed Imager and Spectrometer \citep[NIRI,][]{Hodapp03} on Gemini North.  The data presented in this paper originate from the Gemini program GN-2007B-Q-79, observed in queue mode in the period from UT 2007 July 11 to UT 2007 September 30.  The $H$ narrow-band filter of NIRI at a wavelength of $\lambda=1.57 \mu$m, corresponds to the central wavelength of the H$\alpha\lambda6563\mathrm{\AA}$ emission line at the cluster redshift of $z=1.39$.  Figure~\ref{fig:filter} shows the redshift histogram of known cluster members from \citet{Rosati2009} overplotted with the transmission curve of the $H$ narrow-band filter, projected to show redshift coverage. $H$  broad-band observations at $\lambda=1.65 \mu$m were also obtained to measure the underlying continuum.

We observed one pointing of XMMU2235, centred on the coordinates of the brightest cluster galaxy (BCG). The total integration time was split into single exposures of 120 seconds in the $H$ narrow-band and 30 seconds in the $H$ broad-band filter. To remove cosmic rays and bad pixels, each exposure in a set of 10 exposures was offset from the previous one following a fixed dither pattern. The field of view for each single frame is 120$\times$120 arcseconds with a pixel scale of 0.1162 arcsec/pixel.


In total 302 individual $H$ narrow-band and 115 $H$ broad-band exposures were observed. Some of the broad band images and a substantial part of the narrow band exposures suffered from strong fringing, the known NIRI first-frame-bias pattern, and variations in the sky background that could not be removed by the sky subtraction procedure described below. Only frames with a satisfactory sky subtraction were used to construct the combined final image.
This results in a total number of 212 narrow-band exposures and 110 broad-band exposures, corresponding to 424 minutes ($\sim$7 hours) total narrow-band integration time and 55 minutes total broad-band integration time.

Our field of view has a diameter of $\sim 850$ kpc and therefore covers $\sim 400$ kpc in all directions and up to $\sim 550$ kpc along the diagonal. As already noted by \citet{Lidman2008}, the spatial structure of the cluster is elongated at a roughly 45$^\circ$ along the diagonal of our image. This elongation was also reported by \citet{Rosati2009} studying the extended X-ray emission from the Intra Cluster Medium (ICM).

\subsection{Data Reduction}\label{sec:reduction}

The data were reduced with the Gemini IRAF \footnote[1]{IRAF is distributed by the National Optical Astronomy Observatory, which is operated by the Association of Universities for Research in Astronomy (AURA) under cooperative agreement with the National Science Foundation.  The Gemini IRAF package is distributed by Gemini Observatory, which is operated by AURA.} package following standard data reduction procedures.  All images were prepared for the pipeline with the {\tt niprepare} task which adds certain keywords to the image header. Two sets of flat fields were observed each night in each filter. The corresponding flat field exposures were combined and used to create a bad pixel mask for the science exposures with {\tt niflat}.  The most critical procedure is the sky subtraction which is done with {\tt nisky}. Sky images are produced by combining a set of science exposures after masking and subtracting detected objects. Since the sky in the near infrared is very bright as well as variable on time-scales of a couple minutes, only frames that are taken close enough to each other in time are combined. After trying different configurations and comparing the resulting background noise in the sky subtracted image, we find that the best solution for constructing the sky image for each science frame is to combine the frame itself with the four closest exposures. This corresponds to an exposure time of 10 minutes for each combined sky image.

After applying flat field and bad pixel corrections as well as the sky subtraction with the {\tt nireduce} task, all frames of the same filter were combined into a single final science image. To stack the images with the optimal accuracy, first all images taken at one dither position were combined. This has the advantage that the co-adding does not rely on the astrometry of the single images, where almost no sources that could be used to optimize the co-adding are visible. Since the sources in all images at the same dither position have identical image coordinates they can be directly combined into higher signal-to-noise sub-stacks. These 10 sub-stacks (from the 10 different dither positions) of about 20 exposures each were then combined into the final co-added science image  with {\tt imcoadd}. This procedure worked very well for broad-band frames, while for the narrow-band images small offsets between the images that were supposed to be at identical positions were necessary. Only frames with a good sky-subtraction without residuals of fringing, first-frame-bias pattern or sky variability were stacked, resulting in a number of 212 and 110 combined individual exposures in narrow-band and broad-band, respectively.  The final total exposure times are 55 minutes in the $H$ broad-band filter and 424 minutes in the $H$ narrow-band filter.

\begin{figure*}
\includegraphics[width=\textwidth]{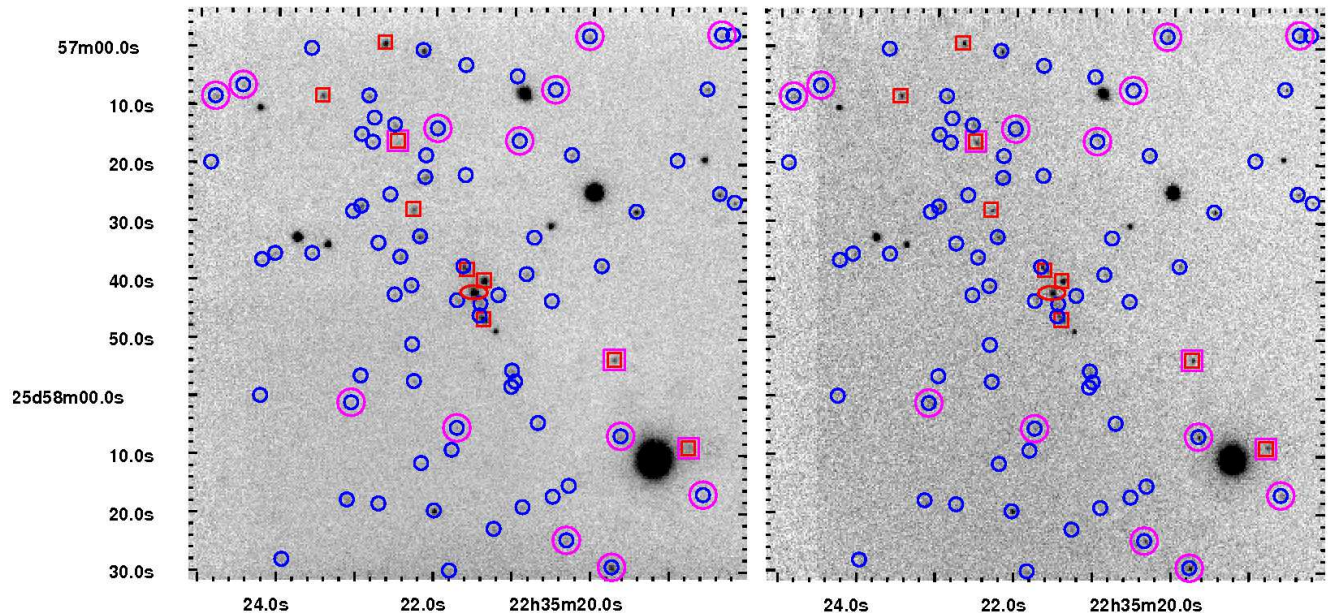}
\caption{Final reduced, combined and trimmed images of XMMU2235. Left panel: $H$ broad-band image. Right panel: $\lambda  1.57 \mu$m  narrow-band image. The width of each image is $\sim1.6$ arcmin or 850 kpc. North is up, east to the left. The spectroscopically confirmed cluster members are marked with red boxes; all other detected objects are marked with blue circles. Objects additionally outlined in magenta are excess emission-line galaxies and most likely to be cluster members (see Figure~\ref{fig:ha_excess}).  The brightest cluster galaxy (BCG), which is also a spectroscopically confirmed member, is marked towards the  centre of each image with a red ellipse.  \label{fig:images}}
\end{figure*}

The combined images in broad and  narrow-band were then trimmed to cover the same field of view in both filters. Figure~\ref{fig:images} shows the two images in the $H$ broad-band (left panel) and narrow-band (right panel).  The images are $1.6$ arcmin $\times$ $1.64$ arcmin, corresponding to about 850 kpc at the cluster's distance. The pixel scale of 0.1162 arcsec/pixel and the angular distance scale at $z=1.39$ of 8.618 kpc/arcsec \citep{Wright2006} give a scale of about 1 kpc/pixel.

\subsection{Source detection}\label{sec:detection}

Sources are detected in the $H$ broad- and narrow-band images using {\tt SExtractor} (Bertin \& Arnouts 1996). To ensure that the different depth of broad- and  narrow-band images does not affect the results, fluxes are measured within the same fixed aperture in both images.  We run {\tt SExtractor} simultaneously on broad- and  narrow-band images, detecting sources in the broad-band image and measuring photometry using the same parameters in both images. The same procedure is repeated using the narrow-band image as the detection image, in order to detect pure emission line sources that might be too faint to be detected in broad-band. The two catalogs are then combined to obtain the final source catalog shown in Table~\ref{tab1}. 
With this method the magnitude of a galaxy in both bands is calculated.  The procedure of detecting sources in both bands ensures that our sample is not biased towards H$\alpha$ emitting galaxies. 

To minimise spurious detections we require a minimum number of 7 adjacent pixels to reside above the background noise by a signal-to-noise ($S/N$) of 2 in the narrow-band image. Obvious non-cluster member detections (e.g. on the edge of the image), stars and known foreground galaxies were removed from the catalog. Stars were excluded based on {\tt SExtractor's CLASS\_STAR} parameter: sources with {\tt CLASS\_STAR} $>$ 0.8 were excluded.  
We use {\tt FLUX\_APER} to measure fluxes within a fixed circular aperture of 20 pixels diameter, corresponding to $2.3^{\prime\prime}$ or $\sim$20 kpc at the cluster distance. 
We compare the use of {\tt FLUX\_APER} and {\tt FLUX\_AUTO} in Section~\ref{sec:aper_compare}.

Our final catalog comprises of 82 objects, shown in Figure~\ref{fig:images} in the broad-band (left) and narrow-band (right), including 10 spectroscopically confirmed cluster members outlined with red boxes.  The spectroscopically confirmed cluster members are taken from \citet{Mullis05}, \citet{Lidman2008}, and \citet{Rosati2009}.  \citet{Rosati2009} find 34 confirmed cluster members from spectra taken with the VLT in Chile covering an area on the sky of $2.0$ arcmin $\times$ $2.0$ arcmin, slightly larger than our coverage but with the same central pointing.  They do not list the coordinates of these objects in the paper.  In fact, none of these papers give object coordinates, but \citet{Mullis05} and \citet{Lidman2008} show images that identify spectroscopically confirmed members.  We compare these published images (Figure~1 in each paper) to our Figure~\ref{fig:images} to identify by eye 10 spectroscopically confirmed cluster members.      

In the subsequent analysis we distinguish spectroscopically confirmed members (referred to as ``confirmed members'') and all other detections by plotting confirmed members in red.  The objects surrounded by an extra magenta symbol in Figure~\ref{fig:images} are galaxies identified as having excess line-emission and are therefore the most likely to be cluster members, as will be discussed in Section~\ref{sec:contaminants}.

The image depth and completeness limits of our broad-band and narrow-band images are estimated by placing artificial point sources in the images with the same noise characteristics as the observed images. The artificial point sources are created assuming a gaussian profile with the width of the point spread function (PSF) measured in the images. The full width at half maximum (FWHM) of the PSF is 6.3 pixels in the broad band image and 4.7 pixels in narrow band image. The same extraction procedure as described above is then run on the images with artificial point sources added. The 5$\sigma$ image depth is given by computing the magnitude of a point source with a flux of five times the background noise within the aperture used for galaxy photometry (10 pixel, $\sim 1$ arcsec radius). We obtain a 5$\sigma$ image depth of $H_{AB} = 24.3$ mag for the broad-band image and $H_{narrow,AB} = 23.3$ mag.  Using our detection criteria, all objects detected in the narrow-band image are also detected in the broad-band image.  
The completeness limit is computed by comparing the number counts per magnitude of the artificial input catalog and the catalog of detected objects in both images. We obtain a 95\% completeness limit of $H_{AB} = 24.2$ mag in the broad-band and $H_{narrow,AB} = 23.4$ mag in the narrow band, values comparable to the 5$\sigma$ image depth.

\subsection{Flux calibration and Star Formation Rate}\label{sec:calibration}

To obtain pure, continuum-subtracted H$\alpha$ fluxes and then star formation rates from the measured fluxes in ADU, we adopted the following procedure: 1. obtain the flux calibration of the  broad-band images, 2. calculate the expected  narrow-band flux from a blackbody SED and relative broad-to- narrow-band throughputs, 3. subtract  broad-band from  narrow-band flux and 4. convert from H$\alpha$ flux (erg~s$^{-1}$) to star formation rates ($M_\odot$yr$^{-1}$).

In order to obtain the AB magnitude zeropoint for the $H$ broad-band images, we observed three standard star fields, taken from the \citet{Persson98} catalog available from the Gemini Observatory website, in both the broad and narrow-band filters.  The standard star counts are measured using {\tt IRAF} and converted to magnitudes applying a correction for airmass and extinction. The  broad-band zeropoint for each star is then computed by adding the measured magnitude to the catalog magnitude. This yields an average  broad-band zeropoint of $ZP_{broad} = 25.43\pm0.03$ magnitudes (AB).

We calculate the expected narrow-band flux of the standard stars using a blackbody SED and the ratio between broad- and  narrow-band throughputs. We assume a blackbody spectrum of the standard stars' respective temperature according to its spectral type. This SED is integrated over the response function plus atmospheric transmission of both broad- and  narrow-band filters. The ratio between expected broad- and  narrow-band flux is then used to compute the expected AB magnitude of the standard stars in the narrow-band. From the expected magnitude and the measured magnitude in the  narrow-band standard star images, we obtain the average  narrow-band zeropoint in AB magnitudes of $ZP_{narrow} = 22.27\pm0.05$ mag. The errors of the two zeropoints are the respective standard deviation of the three measurements of two different stars.

The object magnitudes obtained with the above zeropoints were converted to fluxes using the definition of AB magnitudes \citep{Oke74}: 

\begin{equation}
m(AB) = -2.5 \log(flux) - 48.60,
\end{equation}

\noindent where flux is given in erg~s$^{-1}$ cm$^{-2}$ Hz$^{-1}$. The continuum subtraction is done by subtracting the  broad-band flux from the  narrow-band flux to get the pure H$\alpha$ fluxes. A dust correction of one magnitude is applied to the fluxes \citep{Kennicutt83}, as well as a statistical correction for the contribution of [N~II] emission by a factor of 0.77 \citep{Tresse99}.

The resulting H$\alpha$ fluxes in erg~s$^{-1}$ cm$^{-2}$ Hz$^{-1}$ are then multiplied by the  narrow-band filter width (transformed to Hz) and by the surface of the sphere over which this energy is emitted, i.e. a sphere with the cluster's luminosity distance as radius. We assume that all galaxies are located at the same distance. This then gives the H$\alpha$ luminosities in erg~s$^{-1}$. The errors are based on the zeropoint errors and the measurement errors in counts given by {\tt SExtractor} which are then propagated in the same way as the measurements. We do not take into account any errors in the filter width or the cluster distance.

To convert ergs~s$^{-1}$ to M$_\odot$~yr$^{-1}$ we use the conversion factor of \citet{Kennicutt94}: 
\begin{equation}
1~\textrm{ergs }\,\textrm{s}^{-1}\, = 7.9 \times 10^{-42}~\Msun \,\textrm{yr}^{-1}
\end{equation}
The transformation is known to vary by a factor of $\sim2.5$, introducing an error of $\sim30\%$, which was added to the measurement errors to obtain the final error of the SFRs.  Based on the magnitude limits described above in Section~\ref{sec:detection}, we find a $5~\sigma$ limiting H$\alpha$ luminosity of $L_{H\alpha} = 1\times10^{41}$~erg~s$^{-1}$ which corresponds to a $5~\sigma$ SFR$~=1.1~$M$_\odot$yr$^{-1}$.

\subsection{SFRs measured with different apertures}\label{sec:aper_compare}

When extracting fluxes with {\tt SExtractor}, we test the two different output parameters: {\tt FLUX\_AUTO} and {\tt FLUX\_APER}.  {\tt FLUX\_APER} is the flux measured in a fixed circular aperture of 20 pixels diameter, corresponding to $2.3^{\prime\prime}$ or $\sim$20 kpc at the cluster distance. {\tt FLUX\_AUTO} uses a Kron-like elliptical aperture whose size depends on the objects light distribution. {\tt FLUX\_AUTO} is supposedly more robust \citep{Bertin1996} than {\tt FLUX\_APER} in crowded regions (e.g. the cluster centre). 

The comparison between SFRs measured within the fixed aperture (based on {\tt FLUX\_APER}) and the variable Kron-like aperture (based on {\tt FLUX\_AUTO}) shows that they agree well within the errors. There is only one outliers, which is a confirmed member located very close to the brightest star in the image. This galaxy has a SFR $\sim$ 3.5 ~M$_\odot$yr$^{-1}$ based on the aperture flux, but has a much lower SFR ($\sim$ -2~M$_\odot$yr$^{-1}$) based on {\tt FLUX\_AUTO}. This is probably due to the deblending algorithm of {\tt SExtractor} affecting the detection area that is assigned to galaxies with close neighbours (in this case a bright star). Measuring the flux within a fixed aperture is more robust in these cases, which is reflected in the large measurement error of the {\tt FLUX\_AUTO}-SFR of $\pm$2.2~M$_\odot$yr$^{-1}$, compared to $\pm$0.7~M$_\odot$yr$^{-1}$ for the {\tt FLUX\_APER}-SFR. 
For the whole sample the SFRs measured within a fixed aperture by {\tt FLUX\_APER} show on average slightly lower errors than the SFRs measured with a Kron-like aperture and we therefore adopt the fixed aperture SFRs in the following. 

\subsection{Cluster Membership}\label{sec:contaminants}

\begin{figure}
\includegraphics[width=0.47\textwidth]{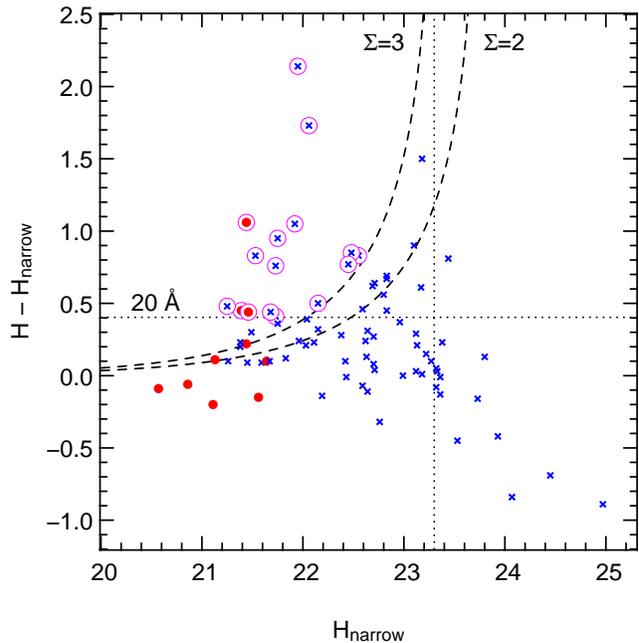}
\caption{Colour-magnitude diagram for all 82 sources detected in the narrow-band image.  The dashed lines indicate lines of constant significance $\Sigma$, which show the excess flux in the narrow-band image above the noise in the broad-band image.  Spectroscopically confirmed cluster member galaxies are plotted as large red points while all other detected objects are shown as blue crosses.  The vertical line shows the 95\% completeness level, so only candidates brighter than this limit are considered as possible excess emission-line objects.  Candidate emission-line cluster galaxies are objects with $EW > 20 \mathrm{\AA}$ and $\Sigma > 3$ (open magenta circles).  \label{fig:ha_excess}}
\end{figure}

We detect 82 resolved objects at the 2$\sigma$ level in the narrow-band and broad-band images.  Of these, 10 are confirmed cluster members with spectroscopic redshifts.  H$\alpha$ emitters at the redshift of the cluster can be identified among the remaining 72 candidate cluster members by having excess flux in the narrow-band relative to the broad-band observations.  

Figure~\ref{fig:ha_excess} shows the excess flux in the narrow-band filter as a function of magnitude for all 82 detected objects.  We require the narrow-band flux to be a factor of three $\Sigma$ greater than the noise in the broad-band (continuum) image \citep{Bunker1995} for an object to be considered an excess line-emitter.  The dotted curves in Figure~\ref{fig:ha_excess} show lines of constant $\Sigma$.  Emission-line candidates with low equivalent widths ($EW$) are likely to be foreground interlopers \citep{Palunas2004}, so we also apply a minimum $EW$ for the H$\alpha$ line of 20$\mathrm{\AA}$ \citep{Hayes2010}.  Candidate emission-line cluster galaxies are objects with $EW > 20 \mathrm{\AA}$ and excess line-emission of $\Sigma > 3$, identified as open magenta circles in Figure~\ref{fig:ha_excess}.  

Of all the detected objects, 16 satisfy the criteria to be considered excess line-emitters and are therefore more likely to be members of the cluster.  The red points in Figure~\ref{fig:ha_excess} show spectroscopically confirmed cluster members, which are all on the bright end ($H_{narrow}<22$) of the distribution of narrow-band magnitudes of detected objects.  Even though these objects are spectroscopically confirmed as cluster members, only three of the 10 satisfy the criteria for being excess H$\alpha$ line-emitters associated with the cluster.  This shows that while this method is a good test to show the most likely cluster candidate members based on excess line emission when no spectroscopic confirmation is available, if a galaxy in the cluster does not have excess line emission (ie. it is not star-forming), it will not be identified as a member of the cluster by this technique.  \citet{Lidman2008} argue that galaxies in the XMMU2235 cluster centre are very likely cluster members based on the infrared colour-magnitude relation of the cluster and the low  likelihood of finding foreground or background galaxies in the narrow colour interval.  
  
What objects could act as possible interlopers?  Galaxies bright in $H\alpha$ could be background galaxies with [OII] emission at $z=3.21$, but this is unlikely due to their generally faint $H$ narrow-band magnitudes.  Galaxies that are much fainter in the narrow-band than expected from their  broad-band magnitude are interpreted to have H$\alpha$ in absorption and therefore show negative SFRs.  The absorption could be caused by stellar absorption of the Balmer line, or a strong non-H$\alpha$ absorption feature in foreground galaxies.  Galaxies at $z\sim0.85$ would have calcium triplet absorption lines at  $\lambda$$\lambda$$\lambda$8498,8542,8662$\mathrm{\AA}$ that would fall into the $\lambda 1.57 \mu$m  narrow-band filter.  Of 15 galaxies that show negative SFRs, the 4 strongest are confirmed cluster members (the 4 bright galaxies in the cluster centre).  
 
Throughout the rest of the paper, we continue to identify confirmed cluster members in red, the candidate members satisfying the excess line-emission criteria in blue, and the remaining object detections in grey.

\section{Results}\label{sec:results}

In the following section we present the calculated star formation rates (SFRs) and how they correlate with properties of the galaxies (magnitude) and the environment (distance from the cluster centre, as defined by the BCG).  Table~\ref{tab1} gives the results for all 82 objects we consider in this study with spectroscopically confirmed cluster members listed as the first 10 items.  The last column gives information about cluster membership and how detected objects are classified (see Section~\ref{sec:detection}):  Excess line-emitters as defined in Section~\ref{sec:contaminants} have a flag corresponding to their $\Sigma$-level (2 or 3), while galaxies below that have a flag $=1$. 

For the full sample of 82 galaxies, we find an average SFR of 1.0~M$_\odot$yr$^{-1}$ with a scatter of 1.6.   For the spectroscopically confirmed cluster members, we measure an average SFR of 1.0$~\pm~2.6~$M$_\odot$yr$^{-1}$.  Among just the excess line-emission galaxies we find an average SFR of 3.6$~\pm~1.3$~M$_\odot$yr$^{-1}$.

\subsection{SFR as a function of $H$ broad-band magnitude}\label{sec:sfr-mag}

\begin{figure}
\includegraphics[width=0.485\textwidth]{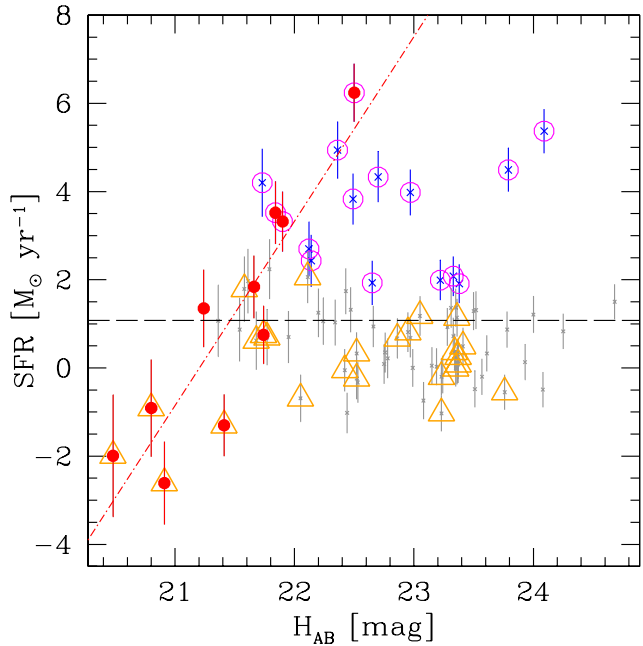}
\caption{Star formation rates as a function of $H$ broad-band magnitude. Spectroscopically confirmed member galaxies are plotted as large red circles; excess line-emission candidate member galaxies are large blue crosses and/or open magenta circles; all other detections are small grey crosses. The red dashed-dotted line represents a least squares fit to the data points of confirmed members. The $5\sigma$ SFR detection limit is shown as the dashed horizontal line, as described in Section~\ref{sec:calibration}. Galaxies within a 200 kpc radius of the BCG are marked with open orange triangles. \label{fig:sfr-hmag}}
\end{figure}

We now present the correlation between SFR and an intrinsic galaxy property, galaxy magnitude. The central wavelength of the observed $H$ broad-band corresponds to a restframe wavelength of $\sim$6900$\mathrm{\AA}$, slightly redder than restframe $R$ band.  Figure~\ref{fig:sfr-hmag} shows the SFRs as a function of $H$ broad-band magnitude.  We find a dependence of SFR on galaxy $H$ broad-band magnitude such that brighter galaxies have lower SFRs.  This correlation is strong for the confirmed members (red) but less evident for excess line-emission galaxies (blue).  A linear fit to the 10 data points of the confirmed members is shown as red dashed-dotted line in Figure~\ref{fig:sfr-hmag}. 

We calculate Spearman's rank correlation coefficients and Kendall's $\tau$ separately for confirmed and excess line-emission galaxies. The correlation coefficient, $\rho$, can have values between $-1 \leq \rho \leq 1$, where $\rho =1$ ($\rho =-1$) means that the two data sets are correlated (anti-correlated) by a monotonic function and $\rho =  0$ indicates that the two data sets are completely uncorrelated. The significance of $\rho$ can be calculated from $\rho$ and the sample size $N$. We give the significance in terms of the percentage likelihood that there is no correlation between the two variables.  Since $\rho$ is less suited for small sample sizes we also compute Kendall's $\tau$ coefficient to support the results on statistical significance for the small subsample of confirmed members ($N=10$). Kendall's $\tau$ is analogous to $\rho$, however, it computes a probability rather than estimating the amount of variation in the data that can be accounted for by a monotonic function. Kendall's $\tau$ is the difference between the two probabilities that the data are observed in the same or in a different order, giving an estimate of the likelihood of a correlation between the variables, or more accurately the lack thereof.

We find that confirmed members have a correlation coefficient of $\rho_{mem} = 0.90$ with a probability of $P_{mem} = 5\%$ that the two variables are uncorrelated. Kendall's $\tau$ evaluates the correlation at a slightly higher confidence with $\tau=0.69$ and a corresponding probability for the lack of a correlation of $P_{mem}=0.7\%$.  The candidate members do not show any significant correlation between SFR and magnitude. We compute $\rho_{cand} = 0.04$ with a probability $P_{cand} = 44\%$ and $\tau=-0.06$ with a probability of $P_{cand} = 43\%$.  The same is true for excess line-emitters, which show no correlation between SFR and $H$ broad-band magnitude ($P_{excess} = 42\%$).  We conclude that for spectroscopically confirmed cluster members there is a positive correlation between SFR and magnitude, i.e. the SFR increases for fainter galaxies in the cluster. 

It is possible that a selection effect exists which causes faint galaxies to be less likely to get a successful redshift determination.  Redshifts from \citet{Rosati2009} were determined from a combination of cross-correlations with a range of template spectra and identification of the [OII]$\lambda3727\mathrm{\AA}$ emission line, although they claim most redshifts were derived solely from the [OII] emission line.  Among galaxies with the same $H$-band magnitude, the galaxy spectra with strong emission lines are more likely to result in successful redshift determinations than the spectra with only weak or no emission.  These galaxies are most likely to be those identified as excess line-emission galaxies shown in blue in Figure~\ref{fig:sfr-hmag}.

\begin{figure}
\includegraphics[width=0.485\textwidth]{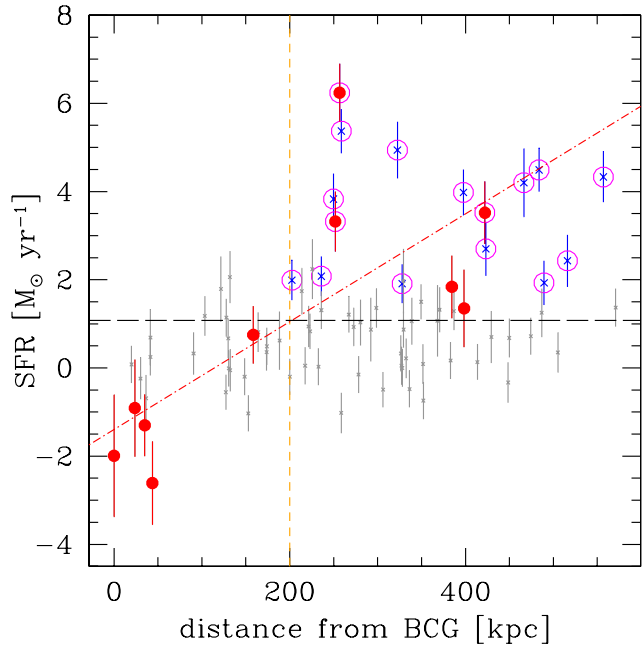}
\caption{Star formation rates as a function of distance from the cluster centre. Symbols are the same as shown in Figure~\ref{fig:sfr-hmag}. The vertical dashed line marks the projected distance within which no star formation occurs ($R_{QUENCH}$). Galaxies within this limit are also marked with orange triangles in Figure~\ref{fig:sfr-hmag}. \label{fig:sfr-dist}}
\end{figure}

\subsection{SFR as a function of distance from the cluster centre}\label{sec:sfr-dist}

In the following we show the correlation between the SFRs and the distance from the brightest cluster galaxy (BCG: see Figure~\ref{fig:images}). \citet{Rosati2009} show that the position of the BCG coincides within 2$^{\prime\prime}$ of the peak of the cluster's extended X-ray emission. It is therefore justified to assume that the BCG resides in the centre of the cluster's potential well and can be regarded as the cluster centre. 

Figure~\ref{fig:sfr-dist} shows the radial distribution of SFRs within the cluster.  While the measured SFRs do not cover a large range in values, we find that SFR increases with distance from the cluster centre.  In fact, there are no star-forming galaxies among the confirmed cluster members and excess line-emission galaxies within a radius of $r\sim 200$ kpc of the BCG.  The five confirmed members within a 200 kpc radius have no excess H$\alpha$ emission.  The confirmed members outside a 200 kpc radius have SFRs of about 1.5~$M_\odot$yr$^{-1}$ or above.  The same is true for all excess line-emission member candidates: there are no star forming candidate cluster members within $r \sim 200$ kpc.  All galaxies with SFR $>$ 3~M$_\odot$yr$^{-1}$ are located at radii larger than $\sim 250$ kpc. 

The SFR-distance relation is significant for member candidates. The Spearman's rank correlation coefficients for members and candidates respectively are $\rho_{mem} = 0.73$ at $P_{mem} = 15\%$ and $\rho_{cand} = 0.42$ at $P_{cand} = 5\%$. Kendall's $\tau$ coefficient gives a likelihood that SFR and distance are uncorrelated of $P_{mem} = 5\%$ and $P_{cand} = 0.1\%$ for members and candidates, respectively.  If we only consider excess line-emitting galaxies we derive a probability of $P_{excess} = 48\%$, concluding that SFR and cluster-centric distance are uncorrelated for the excess line-emitters.
Furthermore, the faint galaxies that are located in the projected cluster centre are too quiescent for their $H$ band magnitude, as suggested from the SFR-magnitude relation presented in Section~\ref{sec:sfr-mag}. Galaxies within a radius of 200 kpc from the BCG are marked with orange triangles in Figure~\ref{fig:sfr-hmag}. The five galaxies that show the largest offset from the relation shown for confirmed cluster members (red dashed-dotted line) are located at the smallest projected distance from the BCG. We suggest that the star formation in those faint galaxies is quenched by processes occurring in the cluster centre. 

Instead of a continuous decrease in star formation we see evidence for a quenching radius $R_{QUENCH}$, within which the star formation is sharply shut off among galaxies confidently associated with the cluster. If galaxies within $R_{QUENCH}\sim 200$ kpc are not considered, the correlation between SFR and cluster-centric distance largely disappears. To distinguish between the two hypotheses of (1) the existence of a quenching radius, as opposed to (2) continuously decreasing SFRs as field galaxies are accreted onto the cluster, $H\alpha$ imaging of a larger field of view out to larger cluster-centric distances is necessary.  This work represents one of the first studies of a $z>1$ galaxy cluster with solid measurements of SFRs of individual galaxies from which we can see the evidence for a SFR-distance relation and possible quenching radius.

\section{Discussion}\label{sec:discussion}

Recent studies of galaxy clusters at intermediate redshift and $z>1$ indicate varying results regarding star-forming trends among galaxies, although most studies show evidence for increasing SFRs with distance from the cluster centre.  We now look more in depth at previous studies.

\citet{Finn05} study three intermediate redshift clusters from the ESO Distant Cluster Survey (EDisCS) in the redshift range $0.65 < z < 0.95$ and compare their median SFRs to field galaxies in the same redshift range. They find that the median SFR of cluster members is between 2 and 3.5 M$_\odot$yr$^{-1}$, which is significantly lower than the median SFR of field galaxies of about 6 M$_\odot$yr$^{-1}$,  showing that star formation in clusters is suppressed at intermediate redshifts.  We measure a median SFR for XMMU2235 of 1.0~M$_\odot$yr$^{-1}$ with a scatter of 1.2 for all 82 galaxies, consistent with the SFRs of EDisCS clusters but nearly a factor of 10 lower than the typical field galaxy SFR value of 10 M$_\odot$yr$^{-1}$ for this redshift range \cite[e.g.,][]{Bauer05,noeske07,Damen09,Bauer2010}.     

Interestingly, our results show that star formation in the central region of a dense cluster at $z=1.4$ is suppressed relative to the field, and there is evidence for the suppression of star formation to not be as strong inside a structure at $z=1.6$ that appears to be a galaxy cluster in formation.  \citet{Kurk2009} identify an over-density at $z=1.6$ called Cl 0332-2742 in the GMASS survey.  Using Spitzer/MIPS observations, they find that SFRs of galaxies within the over-density are only three to five times lower than SFRs of field galaxies at similar redshifts, as opposed to the factor of 10 decrease we find in the XMMU2235 cluster.  The Cl 0332-2742 over-density is most likely not yet virialised, which could explain why galaxies in this structure still show significant levels of star formation. 

Two of the three clusters studied by \citet{Finn05} also show a decrease in H$\alpha$ fraction towards cluster centre.  In contrast to this result, a narrow-band H$\alpha$ study by \citet{Koyama09} of the RXJ1716.4+6708 cluster at $z=0.81$, finds no H$\alpha$ emitters within a radius of 250 kpc of the cluster core.  \citet{Hayashi10} identify 44 [OII] emitters in the central region of the XMMXCS J2215.9-1738 cluster at $z=1.46$ and calculate SFRs down to 2.6~M$_\odot$yr$^{-1}$.  They find no evidence for decreasing SFRs towards the cluster core.  

Our findings agree with a recent study of SFRs in the cluster RX J0152.7-1357 at $z=0.834$ by \citet{Patel2009}.  They stack Spitzer/MIPS imaging at 24 $\mu$m to determine SFRs and find a decrease in SFR with increasing local density within a similar density regime to our cluster.  Both this study and that of \citet{Patel2009} contrast to recent studies of field galaxies at $z\sim1$ by \citet{Elbaz07} and \citet{Cooper08} who suggest that at $z\sim1$, galaxies in higher density environments exhibit higher SFRs than galaxies at lower densities.  The discrepancy is likely due to differing definitions of environment and local density \citep[see discussion in][]{Patel2009}.  \citet{Cooper08} examine galaxies in the DEEP2 survey, while \citet{Elbaz07} study the GOODS fields, neither of which contains galaxy clusters.  The decrease of SFR we find at the highest densities, meaning the cluster centre, is not a density region probed by the studies of \citet{Elbaz07} or \citet{Cooper08} which might be the reason that they do  not observe the suppressed SF at high densities which we see in XMMU2335.

Other studies of the XMMU2235 cluster include \citet{Lidman2008}, \citet{Rosati2009}, and very recently, \citet{Strazzullo2010}.  \citet{Lidman2008} use simple stellar population models to determine that galaxies in the core of XMMU2235 are already old, with typical ages of 3 Gyr.  \citet{Rosati2009} detect no [OII] emitters within 250 kpc of the cluster centre, while \citet{Strazzullo2010} use rest-frame ultraviolet properties of spectroscopically confirmed members to show that individual galaxies within a clustercentric distance of 250 kpc are effectively quenched of star formation.  These results are consistent with our findings of no new star formation in the core, including evidence for the presence of H$\alpha$ absorption in the four confirmed cluster members located within a radius of $\sim 100$~kpc, and a larger scatter in SFR values with increasing distance from the central cD galaxy.   We confirm evidence for the build-up of the red sequence starting in the cluster core and moving outwards.  Furthermore we find that there is a radius, $R_{QUENCH}$ of about 200 kpc, within which star formation is efficiently quenched. 

Environmental effects that shut down star formation in cluster galaxies are generally assumed to begin when accreting satellite galaxies have reached the virial radius of a host halo.  However, some studies have shown that SFRs can remain depressed relative to the field out to two or three times the virial radius of the cluster \citep{Balough2000,Verdugo2008}.  The virial radius of  XMMU2235 is estimated to be around 1 Mpc \citep{Rosati2009,Jee2009}, which is well beyond the radius we associate with complete star formation quenching, $R_{QUENCH}$, and much larger than our observations extend.  Note that the $R_{QUENCH}$ we define can be thought of as a lower limit due to projection effects.  It is possible that SFRs outside of $R_{QUENCH}$ remain around the low values we measure until the virial radius of the cluster, or farther, and then increase to match the value of field galaxies at that redshift.     

We interpret the possible existence of $R_{QUENCH}$ as evidence for suppressed star formation due to cluster specific processes at the very central region of the cluster.  Gas could be removed from galaxies through ram pressure stripping as they enter the cluster environment and pass through the dense intergalactic medium in the cluster centre \citep[e.g.][]{gunn-gott1972,Bekki2009,Tonnesen2010}.  A galaxy could also lose its halo gas reservoir according to hierarchical formation models, in a process called ``strangulation'' \citep{Larson1980}.  Such processes have recently been modelled by \citet{Book2010}, but they were unable to determine the exact causes of gas-stripping in cluster environments due to the ``complex interplay of preprocessing effects at work.''   Another process causing the suppression of star formation is galaxy harassment, repeated high velocity encounters and interactions, which require both high relative velocities and high galaxy number densities; conditions that are present in the centre of massive galaxy clusters like XMMU2235.   In fact, the velocity dispersion of XMMU2235 is  $\sim800$~km/s \citep{Rosati2009} which means the crossing time at $R_{QUENCH}$ is quite short, at roughly $0.25$~Gyr.

In this study we probe relatively small cluster-centric radii and (maybe consequently) measure a relatively narrow range of SFRs up to $\sim$ 6 M$_\odot$yr$^{-1}$. To investigate if the SFRs in the cluster outskirts reach the typical field value at this redshift of 10 M$_\odot$yr$^{-1}$ or if they are undergoing a short initial increase in SFR before quenching sets in, observations of individual galaxies  across a larger distance from the cluster centre are needed.

\section{Summary and conclusions}\label{sec:summ}

We examine star formation rates (SFRs) in the high redshift, massive galaxy cluster XMMU2235 at $z=1.39$ \citep{Mullis05}.  To calculate SFRs we use deep $H$ narrow-band imaging from NIRI at Gemini North, which corresponds to the wavelength of the H$\alpha$ emission line at the redshift of the cluster.  XMMU2235 is the most massive galaxy cluster above $z\sim1$ known to date \citep{Rosati2009} and has a population of red, supposedly passive galaxies forming a relatively tight red sequence \citep{Lidman2008}. Our observed field is 1.6$\times$1.6 arcminutes which corresponds to roughly 500$\times$500 kpc.  In the following we summarize our most important findings:

\begin{enumerate}
\renewcommand{\theenumi}{\arabic{enumi}.}

\item  We directly measure the excess H$\alpha$ line emission of 82 individual objects inside the observed field of the massive galaxy cluster, XMMU J2235.3-2557 at $z=1.39$.  Of the 82 galaxies, 10 galaxies are spectroscopically confirmed cluster members. A total of 16 out of the 82 galaxies are identified as excess line-emission galaxies and therefore the most likely to be candidate cluster members.  We estimate SFRs for all detected objects from the the narrow-band imaging of H$\alpha$ at $z=1.39$.  \\

\item We find a small range in SFRs from zero to 6 M$_\odot$yr$^{-1}$ within the surveyed area of $\sim500$ kpc radius.   Including all observed galaxies, we measure a median SFR of 1.0~M$_\odot$yr$^{-1}$ with a scatter of 1.6, consistent with the SFRs found for intermediate redshift clusters (e.g. EDisCS clusters at $0.4<z<0.8$) and a factor of $\sim$10 lower than field galaxies at $1<z<2$.  Among just the excess line-emission galaxies we find an average SFR of 3.6$~\pm~1.3$~M$_\odot$yr$^{-1}$.  The highest SFRs in the cluster are $\sim$~6~M$_\odot$yr$^{-1}$, still lower than the median SFR of field galaxies at this redshift.\\  

\item  SFRs increase with distance from the brightest cluster galaxy.  All excess line-emitting galaxies are located outside a radius of $\sim200$ kpc from the brightest cluster galaxy.  All galaxies with SFR~$>$~3~M$_\odot$yr$^{-1}$ are located outside a radius of $\sim250$~kpc from the cluster centre.  We argue that the star formation might not only be continuously suppressed as galaxies approach the cluster centre, but that there is a quenching radius ($R_{QUENCH}~\sim~200$~kpc) within which star formation is rapidly shut-off.
\\  

\item  The lack of star formation within $R_{QUENCH}$ indicates that star formation is suppressed in this massive galaxies cluster at $z=1.39$.  The suppression of star formation could be caused either by interaction with the ICM (ram pressure stripping) or frequent high velocity encounters (harassment) caused by the high galaxy density in the cluster centre where the crossing time at $R_{QUENCH}$ is 0.25 Gyr.

\end{enumerate}

We conclude that star formation in this cluster is effectively shut off in the cluster centre already at $z=1.39$, when the universe was only $\sim$4.5 Gyr old. Galaxies at larger radii from the cluster centre are moderately forming stars, but not reaching the average SFR of field galaxies at this redshift. This suggests that the red sequence is built up from the inside out, starting at redshifts $z > 2$ in the most massive galaxy clusters, in corroboration with previous studies of XMMU2235.  It would be very interesting to test this scenario by proving the SFR-distance relation and average SFRs out to larger cluster-centric distances.

\section*{Acknowledgments}

Based on observations obtained at the Gemini Observatory, which is operated by the Association of Universities for Research in Astronomy, Inc., under a cooperative agreement with the NSF on behalf of the Gemini partnership: the National Science Foundation (United States), the Science and Technology Facilities Council (United Kingdom), the National Research Council (Canada), CONICYT (Chile), the Australian Research Council (Australia), Minist\'erio da Ci\^encia e Tecnologia (Brazil) and Ministerio de Ciencia, Tecnolog\'ia e Innovaci\'on Productiva (Argentina).  The data presented in this paper originate from the Gemini program GN-2007B-Q-79, observed in queue mode.  J.V. acknowledges the School of Physics and Astronomy (University of
Nottingham) for granting access to its facilities, as well as all
the members of the School for sharing with him a wonderful working and
personal environment.  

\bibliographystyle{mn2e}

\label{lastpage}

\begin{table*}
\caption{Properties of all objects detected in the H$_{narrow}$ image of the central $\sim500$kpc of the massive galaxy cluster XMMU J2235.3-2557 at $z=1.39$. \label{tab1}}
\begin{tiny}
\begin{tabular}{lcccccccc}
\hline
ID & $\alpha$ & $\delta$ & $H_{broad}$ & $H_{narrow}$ & $L_{H\alpha}$ & SFR & dist. BCG & flag$^1$ \\
  & [hh mm ss] & [dd mm ss] & {\tt MAG\_APER} & {\tt MAG\_APER} & [$10^{40}$ erg~s$^{-1}$] & [$M_\odot$yr$^{-1}$] & [kpc] & \\ 
\hline
1	&	22 35 21.5	&	-25 57 42.2	&	20.48	$\pm$	0.03	&	20.57	$\pm$	0.06	&	-25	$\pm$	18	&	-2.0	$\pm$	1.4	&	0	 & 	1	\\
2	&	22 35 21.4	&	-25 57 40.2	&	20.80	$\pm$	0.04	&	20.86	$\pm$	0.06	&	-11	$\pm$	14	&	-0.9	$\pm$	1.1	&	24	 & 	1	\\
3	&	22 35 21.6	&	-25 57 38.2	&	21.41	$\pm$	0.04	&	21.56	$\pm$	0.07	&	-16	$\pm$	9	&	-1.3	$\pm$	0.7	&	35	 & 	1	\\
4	&	22 35 21.4	&	-25 57 46.7	&	20.91	$\pm$	0.04	&	21.11	$\pm$	0.06	&	-33	$\pm$	12	&	-2.6	$\pm$	0.9	&	44	 & 	1	\\
5	&	22 35 22.3	&	-25 57 28.0	&	21.74	$\pm$	0.05	&	21.64	$\pm$	0.07	&	9	$\pm$	8	&	0.7	$\pm$	0.6	&	159	 & 	1	\\
6	&	22 35 19.7	&	-25 57 53.7	&	21.90	$\pm$	0.05	&	21.46	$\pm$	0.07	&	42	$\pm$	9	&	3.3	$\pm$	0.7	&	252	 & 	3	\\
7	&	22 35 22.5	&	-25 57 16.3	&	22.50	$\pm$	0.07	&	21.44	$\pm$	0.07	&	79	$\pm$	8	&	6.2	$\pm$	0.7	&	257	 & 	3	\\
8	&	22 35 23.5	&	-25 57 08.5	&	21.66	$\pm$	0.05	&	21.44	$\pm$	0.07	&	23	$\pm$	9	&	1.8	$\pm$	0.7	&	384	 & 	1	\\
9	&	22 35 22.7	&	-25 56 59.4	&	21.24	$\pm$	0.04	&	21.13	$\pm$	0.06	&	17	$\pm$	11	&	1.3	$\pm$	0.9	&	398	 & 	1	\\
10	&	22 35 18.8	&	-25 58 08.8	&	21.84	$\pm$	0.05	&	21.39	$\pm$	0.07	&	45	$\pm$	9	&	3.5	$\pm$	0.7	&	422	 & 	3	\\
 \hline               
11 & 22 35 21.4 & -25 57 44.2 & 23.37 $\pm$ 0.12 & 23.32 $\pm$ 0.23 & 1 $\pm$ 5 & 0.1 $\pm$ 0.4 & 20 & 1 \\
12 & 22 35 21.7 & -25 57 43.6 & 22.52 $\pm$ 0.07 & 22.59 $\pm$ 0.13 & -3 $\pm$ 6 & -0.2 $\pm$ 0.5 & 30 & 1 \\
13 & 22 35 21.5 & -25 57 46.1 & 22.05 $\pm$ 0.06 & 22.19 $\pm$ 0.10 & -9 $\pm$ 7 & -0.7 $\pm$ 0.5 & 36 & 1 \\
14 & 22 35 21.7 & -25 57 37.7 & 21.77 $\pm$ 0.05 & 21.67 $\pm$ 0.07 & 9 $\pm$ 8 & 0.7 $\pm$ 0.6 & 41 & 1 \\
15 & 22 35 21.2 & -25 57 42.6 & 23.37 $\pm$ 0.12 & 23.22 $\pm$ 0.21 & 3 $\pm$ 5 & 0.3 $\pm$ 0.4 & 41 & 1 \\
16 & 22 35 20.9 & -25 57 39.1 & 22.52 $\pm$ 0.07 & 22.42 $\pm$ 0.11 & 4 $\pm$ 6 & 0.3 $\pm$ 0.5 & 91 & 1 \\
17 & 22 35 22.3 & -25 57 41.1 & 23.05 $\pm$ 0.10 & 22.59 $\pm$ 0.13 & 15 $\pm$ 6 & 1.2 $\pm$ 0.4 & 103 & 2 \\
18 & 22 35 22.2 & -25 57 32.7 & 21.58 $\pm$ 0.05 & 21.38 $\pm$ 0.07 & 23 $\pm$ 9 & 1.8 $\pm$ 0.7 & 122 & 1 \\
19 & 22 35 22.3 & -25 57 51.2 & 23.76 $\pm$ 0.16 & 24.45 $\pm$ 0.63 & -7 $\pm$ 5 & -0.5 $\pm$ 0.4 & 127 & 1 \\
20 & 22 35 20.8 & -25 57 32.8 & 23.36 $\pm$ 0.12 & 22.80 $\pm$ 0.15 & 14 $\pm$ 5 & 1.1 $\pm$ 0.4 & 128 & 2 \\
21 & 22 35 20.5 & -25 57 43.6 & 22.86 $\pm$ 0.09 & 22.62 $\pm$ 0.13 & 9 $\pm$ 6 & 0.7 $\pm$ 0.5 & 130 & 1 \\
22 & 22 35 22.5 & -25 57 42.7 & 23.35 $\pm$ 0.12 & 23.36 $\pm$ 0.24 & 0 $\pm$ 5 & 0.0 $\pm$ 0.4 & 131 & 1 \\
23 & 22 35 22.5 & -25 57 36.2 & 22.42 $\pm$ 0.07 & 22.43 $\pm$ 0.11 & -1 $\pm$ 6 & -0.1 $\pm$ 0.5 & 132 & 1 \\
24 & 22 35 21.0 & -25 57 55.7 & 22.11 $\pm$ 0.06 & 21.75 $\pm$ 0.08 & 26 $\pm$ 7 & 2.1 $\pm$ 0.6 & 132 & 1 \\
25 & 22 35 21.0 & -25 57 57.5 & 23.23 $\pm$ 0.11 & 23.36 $\pm$ 0.24 & -3 $\pm$ 5 & -0.2 $\pm$ 0.4 & 149 & 1 \\
26 & 22 35 21.0 & -25 57 58.4 & 23.23 $\pm$ 0.11 & 24.07 $\pm$ 0.45 & -13 $\pm$ 5 & -1.0 $\pm$ 0.4 & 153 & 1 \\
27 & 22 35 22.3 & -25 57 57.5 & 22.95 $\pm$ 0.09 & 22.64 $\pm$ 0.13 & 10 $\pm$ 6 & 0.8 $\pm$ 0.4 & 164 & 1 \\
28 & 22 35 21.6 & -25 57 22.1 & 23.41 $\pm$ 0.12 & 23.12 $\pm$ 0.19 & 6 $\pm$ 5 & 0.5 $\pm$ 0.4 & 174 & 1 \\
29 & 22 35 22.8 & -25 57 33.8 & 23.34 $\pm$ 0.12 & 23.13 $\pm$ 0.19 & 5 $\pm$ 5 & 0.4 $\pm$ 0.4 & 174 & 1 \\
30 & 22 35 22.2 & -25 57 22.5 & 21.68 $\pm$ 0.05 & 21.59 $\pm$ 0.07 & 8 $\pm$ 8 & 0.6 $\pm$ 0.7 & 188 & 1 \\
31 & 22 35 22.6 & -25 57 25.5 & 23.57 $\pm$ 0.14 & 23.73 $\pm$ 0.33 & -2 $\pm$ 5 & -0.2 $\pm$ 0.4 & 200 & 1 \\
32 & 22 35 21.7 & -25 58 05.5 & 23.22 $\pm$ 0.11 & 22.45 $\pm$ 0.11 & 25 $\pm$ 6 & 2.0 $\pm$ 0.5 & 203 & 3 \\
33 & 22 35 19.9 & -25 57 37.6 & 22.43 $\pm$ 0.07 & 22.04 $\pm$ 0.09 & 22 $\pm$ 7 & 1.7 $\pm$ 0.5 & 214 & 1 \\
34 & 22 35 22.2 & -25 57 18.8 & 23.15 $\pm$ 0.10 & 23.12 $\pm$ 0.19 & 1 $\pm$ 5 & 0.1 $\pm$ 0.4 & 218 & 1 \\
35 & 22 35 20.7 & -25 58 04.5 & 22.66 $\pm$ 0.08 & 22.38 $\pm$ 0.11 & 12 $\pm$ 6 & 0.9 $\pm$ 0.5 & 221 & 1 \\
36 & 22 35 23.0 & -25 57 56.6 & 24.25 $\pm$ 0.23 & 23.44 $\pm$ 0.25 & 11 $\pm$ 5 & 0.8 $\pm$ 0.4 & 223 & 1 \\
37 & 22 35 23.0 & -25 57 27.5 & 21.79 $\pm$ 0.05 & 21.49 $\pm$ 0.07 & 28 $\pm$ 9 & 2.2 $\pm$ 0.7 & 226 & 1 \\
38 & 22 35 23.1 & -25 57 28.4 & 23.19 $\pm$ 0.11 & 23.18 $\pm$ 0.20 & 0 $\pm$ 5 & 0.0 $\pm$ 0.4 & 233 & 1 \\
39 & 22 35 21.0 & -25 57 16.2 & 23.33 $\pm$ 0.12 & 22.48 $\pm$ 0.12 & 26 $\pm$ 6 & 2.1 $\pm$ 0.4 & 236 & 3 \\
40 & 22 35 21.8 & -25 58 09.3 & 23.52 $\pm$ 0.13 & 22.83 $\pm$ 0.15 & 17 $\pm$ 5 & 1.3 $\pm$ 0.4 & 236 & 2 \\
41 & 22 35 22.0 & -25 57 14.1 & 22.49 $\pm$ 0.07 & 21.73 $\pm$ 0.07 & 48 $\pm$ 7 & 3.8 $\pm$ 0.6 & 250 & 3 \\
42 & 22 35 20.3 & -25 57 18.6 & 22.44 $\pm$ 0.07 & 22.76 $\pm$ 0.14 & -13 $\pm$ 6 & -1.0 $\pm$ 0.5 & 259 & 1 \\
43 & 22 35 23.1 & -25 58 01.2 & 24.09 $\pm$ 0.21 & 21.95 $\pm$ 0.08 & 68 $\pm$ 6 & 5.4 $\pm$ 0.5 & 259 & 3 \\
44 & 22 35 22.2 & -25 58 11.5 & 24.00 $\pm$ 0.19 & 23.10 $\pm$ 0.19 & 15 $\pm$ 5 & 1.2 $\pm$ 0.4 & 267 & 2 \\
45 & 22 35 23.6 & -25 57 35.6 & 23.28 $\pm$ 0.11 & 22.83 $\pm$ 0.15 & 12 $\pm$ 5 & 0.9 $\pm$ 0.4 & 273 & 1 \\
46 & 22 35 22.8 & -25 57 16.5 & 23.24 $\pm$ 0.11 & 23.32 $\pm$ 0.23 & -2 $\pm$ 5 & -0.1 $\pm$ 0.4 & 278 & 1 \\
47 & 22 35 22.6 & -25 57 13.5 & 22.34 $\pm$ 0.06 & 22.11 $\pm$ 0.09 & 13 $\pm$ 7 & 1.0 $\pm$ 0.5 & 281 & 1 \\
48 & 22 35 19.5 & -25 57 28.3 & 21.54 $\pm$ 0.05 & 21.45 $\pm$ 0.07 & 11 $\pm$ 9 & 0.9 $\pm$ 0.7 & 292 & 1 \\
49 & 22 35 23.0 & -25 57 15.2 & 23.31 $\pm$ 0.12 & 22.69 $\pm$ 0.14 & 17 $\pm$ 6 & 1.4 $\pm$ 0.4 & 299 & 2 \\
50 & 22 35 22.8 & -25 57 12.4 & 24.08 $\pm$ 0.20 & 24.97 $\pm$ 1.02 & -6 $\pm$ 5 & -0.5 $\pm$ 0.4 & 306 & 1 \\
51 & 22 35 19.6 & -25 58 06.8 & 22.36 $\pm$ 0.07 & 21.53 $\pm$ 0.07 & 63 $\pm$ 8 & 4.9 $\pm$ 0.6 & 323 & 3 \\
52 & 22 35 21.0 & -25 57 05.1 & 23.36 $\pm$ 0.12 & 23.33 $\pm$ 0.23 & 1 $\pm$ 5 & 0.0 $\pm$ 0.4 & 327 & 1 \\
53 & 22 35 20.3 & -25 58 15.4 & 23.61 $\pm$ 0.14 & 23.38 $\pm$ 0.24 & 4 $\pm$ 5 & 0.3 $\pm$ 0.4 & 327 & 1 \\
54 & 22 35 20.5 & -25 57 07.4 & 23.38 $\pm$ 0.12 & 22.55 $\pm$ 0.12 & 24 $\pm$ 6 & 1.9 $\pm$ 0.4 & 328 & 3 \\
55 & 22 35 20.9 & -25 58 19.1 & 22.99 $\pm$ 0.09 & 22.99 $\pm$ 0.17 & 0 $\pm$ 5 & 0.0 $\pm$ 0.4 & 328 & 1 \\
56 & 22 35 22.0 & -25 58 19.7 & 21.61 $\pm$ 0.05 & 21.38 $\pm$ 0.07 & 25 $\pm$ 9 & 2.0 $\pm$ 0.7 & 329 & 1 \\
57 & 22 35 20.5 & -25 58 17.3 & 23.78 $\pm$ 0.16 & 23.17 $\pm$ 0.20 & 11 $\pm$ 5 & 0.9 $\pm$ 0.4 & 330 & 1 \\
58 & 22 35 24.1 & -25 57 35.7 & 22.78 $\pm$ 0.08 & 22.70 $\pm$ 0.14 & 3 $\pm$ 6 & 0.2 $\pm$ 0.4 & 332 & 1 \\
59 & 22 35 21.7 & -25 57 03.3 & 23.51 $\pm$ 0.13 & 23.93 $\pm$ 0.39 & -6 $\pm$ 5 & -0.5 $\pm$ 0.4 & 336 & 1 \\
60 & 22 35 22.9 & -25 57 08.6 & 22.24 $\pm$ 0.06 & 22.03 $\pm$ 0.09 & 13 $\pm$ 7 & 1.1 $\pm$ 0.5 & 339 & 1 \\
61 & 22 35 22.7 & -25 58 18.5 & 24.68 $\pm$ 0.33 & 23.18 $\pm$ 0.20 & 19 $\pm$ 5 & 1.5 $\pm$ 0.4 & 350 & 2 \\
62 & 22 35 24.2 & -25 57 36.8 & 23.08 $\pm$ 0.10 & 23.53 $\pm$ 0.27 & -9 $\pm$ 5 & -0.7 $\pm$ 0.4 & 352 & 1 \\
63 & 22 35 21.3 & -25 58 22.8 & 22.75 $\pm$ 0.08 & 22.71 $\pm$ 0.14 & 1 $\pm$ 6 & 0.1 $\pm$ 0.5 & 352 & 1 \\
64 & 22 35 22.2 & -25 57 00.7 & 21.36 $\pm$ 0.04 & 21.26 $\pm$ 0.06 & 13 $\pm$ 10 & 1.1 $\pm$ 0.8 & 368 & 1 \\
65 & 22 35 23.1 & -25 58 17.9 & 22.47 $\pm$ 0.07 & 22.15 $\pm$ 0.09 & 17 $\pm$ 6 & 1.3 $\pm$ 0.5 & 371 & 1 \\
66 & 22 35 24.2 & -25 58 00.1 & 23.37 $\pm$ 0.12 & 23.27 $\pm$ 0.22 & 2 $\pm$ 5 & 0.2 $\pm$ 0.4 & 383 & 1 \\
67 & 22 35 19.0 & -25 57 19.5 & 23.50 $\pm$ 0.13 & 22.83 $\pm$ 0.15 & 16 $\pm$ 5 & 1.3 $\pm$ 0.4 & 387 & 2 \\
68 & 22 35 20.3 & -25 58 24.7 & 22.97 $\pm$ 0.09 & 21.92 $\pm$ 0.08 & 50 $\pm$ 7 & 4.0 $\pm$ 0.5 & 398 & 3 \\
69 & 22 35 21.8 & -25 58 30.0 & 23.93 $\pm$ 0.18 & 23.80 $\pm$ 0.35 & 2 $\pm$ 5 & 0.1 $\pm$ 0.4 & 414 & 1 \\
70 & 22 35 20.1 & -25 56 58.2 & 22.12 $\pm$ 0.06 & 21.68 $\pm$ 0.07 & 34 $\pm$ 8 & 2.7 $\pm$ 0.6 & 423 & 3 \\
71 & 22 35 18.4 & -25 57 25.1 & 21.95 $\pm$ 0.05 & 21.83 $\pm$ 0.08 & 9 $\pm$ 7 & 0.7 $\pm$ 0.6 & 429 & 1 \\
72 & 22 35 18.2 & -25 57 26.6 & 22.53 $\pm$ 0.07 & 22.64 $\pm$ 0.13 & -4 $\pm$ 6 & -0.3 $\pm$ 0.5 & 449 & 1 \\
73 & 22 35 23.6 & -25 57 00.5 & 22.97 $\pm$ 0.09 & 22.70 $\pm$ 0.14 & 9 $\pm$ 6 & 0.7 $\pm$ 0.4 & 450 & 1 \\
74 & 22 35 19.8 & -25 58 29.3 & 21.73 $\pm$ 0.05 & 21.25 $\pm$ 0.06 & 53 $\pm$ 10 & 4.2 $\pm$ 0.8 & 466 & 3 \\
75 & 22 35 24.9 & -25 57 20.1 & 23.33 $\pm$ 0.12 & 22.96 $\pm$ 0.17 & 9 $\pm$ 5 & 0.7 $\pm$ 0.4 & 474 & 1 \\
76 & 22 35 18.6 & -25 58 16.7 & 23.79 $\pm$ 0.16 & 22.06 $\pm$ 0.09 & 57 $\pm$ 6 & 4.5 $\pm$ 0.5 & 484 & 3 \\
77 & 22 35 18.6 & -25 57 07.2 & 22.20 $\pm$ 0.06 & 21.96 $\pm$ 0.08 & 16 $\pm$ 7 & 1.2 $\pm$ 0.5 & 487 & 1 \\
78 & 22 35 24.5 & -25 57 06.8 & 22.65 $\pm$ 0.08 & 22.15 $\pm$ 0.09 & 24 $\pm$ 6 & 1.9 $\pm$ 0.5 & 489 & 3 \\
79 & 22 35 24.0 & -25 58 28.2 & 22.76 $\pm$ 0.08 & 22.63 $\pm$ 0.13 & 4 $\pm$ 6 & 0.4 $\pm$ 0.5 & 505 & 1 \\
80 & 22 35 24.8 & -25 57 08.7 & 22.14 $\pm$ 0.06 & 21.73 $\pm$ 0.07 & 31 $\pm$ 8 & 2.4 $\pm$ 0.6 & 516 & 3 \\
81 & 22 35 18.4 & -25 56 57.8 & 22.70 $\pm$ 0.08 & 21.75 $\pm$ 0.08 & 55 $\pm$ 7 & 4.3 $\pm$ 0.6 & 557 & 3 \\
82 & 22 35 18.3 & -25 56 57.7 & 23.35 $\pm$ 0.12 & 22.71 $\pm$ 0.14 & 17 $\pm$ 5 & 1.4 $\pm$ 0.4 & 571 & 2 \\
\hline
\end{tabular}
\begin{flushleft}
$^1$ Spectroscopically confirmed members are the first 10 objects. Galaxies that are excess line-emitters have a flag of 2 (2$\sigma$ confidence) and 3 (3$\sigma$ confidence). 
\end{flushleft}
\end{tiny}
\end{table*}

\end{document}